\documentclass{article}

\usepackage{arxiv}

\usepackage{graphicx}
\usepackage{subcaption}
\usepackage{svg} % allow inclusion of .svg images
\usepackage{xcolor}
\graphicspath{ {./images/} }

\usepackage{hyperref}
\usepackage{url}
\usepackage[usenames,dvipsnames]{pstricks}
\usepackage{epsfig}
\usepackage{pst-grad} % For gradients
\usepackage{pst-plot} % For axes
\usepackage{amsmath,epsfig}
\usepackage{amssymb}
\usepackage{psfrag}
\usepackage{color}
\usepackage{pstricks}
\usepackage{cite}
\usepackage{algorithm}
\usepackage{epsfig}
\usepackage{amsthm}
\usepackage{amssymb}
\usepackage{psfrag}
\usepackage{pstricks}
\usepackage{float}
\usepackage{stfloats}
\usepackage{wrapfig}
\usepackage{graphicx}
\usepackage{amsmath}
\usepackage{amsfonts}
\usepackage{amssymb}
\usepackage{blindtext}
\usepackage{array}
\usepackage{arydshln}
\usepackage{pstricks,pst-plot}
\usepackage{pst-bezier}
\usepackage{pst-math}
\usepackage{algorithm,algcompatible,amsmath}
\usepackage{float}
\usepackage[mathscr]{eucal}
\usepackage{enumerate}
\usepackage{hyperref}
\usepackage{multirow}
\usepackage{graphicx}
\usepackage{psfrag}
\usepackage{pstool}
\usepackage{tikz}
\usepackage{booktabs}

\usetikzlibrary{arrows.meta}

\def\b0{{\bf 0}}

\title{Characterizing Parallelism Strategies in LLM Inference: Fundamental Compute-Communication Trade-offs }

\author{
 \large Javad Mirzaei\thanks{Corresponding author.} \\
   Office of CTO\\
  Dell Technologies Inc. \\
  {Canada} \\
  \texttt{javad.mirzaei@dell.com} 
   \And
  \large Jeebak Mitra \\
  Office of CTO\\
  Dell Technologies Inc. \\
  {Canada} \\
  \texttt{jeebak.mitra@dell.com}
}

\begin{document}
\maketitle
\begin{abstract}
Large Language Model (LLM) inference has become the dominant workload in modern AI systems, requiring serving infrastructures to simultaneously maximize throughput while satisfying stringent latency Service-Level Objectives (SLOs). Since state-of-the-art LLMs exceed the compute and memory capacity of a single GPU, inference is commonly distributed across multiple GPUs using tensor parallelism (TP), pipeline parallelism (PP), or hybrid parallelism (HB), which combines both approaches. However, selecting an appropriate parallelism strategy remains a challenging task because inference performance is governed by complex interactions among computation, inter-GPU communication, pipeline utilization, sequence length, batch size, and model architecture. Existing studies primarily rely on empirical evaluations or optimize specific serving components, offering limited analytical insights into the performance trade-offs among different parallelism strategies, particularly across the distinct prefill and decoding phases of inference.

In this paper, we present a unified analytical framework for modeling distributed LLM inference under TP, PP, and HB. Our framework decomposes end-to-end inference latency into computation, inter-GPU communication, and pipeline bubble overhead, and derives analytical models that explicitly characterize TP collective communication, PP point-to-point communication, and pipeline utilization as functions of hardware, model and workload characteristics. The proposed model further captures the fundamentally different execution characteristics of prefill and decoding, explaining why PP-dominant configurations favor compute-intensive prefill while TP configurations achieve lower decoding latency by eliminating pipeline bubbles. Extensive experiments using modern LLMs on multi-GPU platforms validate the model and confirm the fundamental trade-off between compute and communication of TP and PP across diverse workloads. Our framework provides practical insights for selecting efficient parallelism strategies and establishes a principled foundation for performance modeling and capacity planning in future LLM serving systems.
\end{abstract}
\section{Introduction}

Large Language Models (LLMs) have rapidly become a fundamental component of modern artificial intelligence (AI) applications, enabling conversational assistants, retrieval-augmented generation, code generation, and long-context reasoning systems. The rapid evolution of models from billions to hundreds of billions of parameters, together with the increasing adoption of long-context workloads, has introduced unprecedented computational and memory challenges during inference. Unlike model training, where large batches and long-running executions allow systems to maximize accelerator utilization, LLM serving must simultaneously achieve high throughput and satisfy strict latency requirements, including time-to-first-token (TTFT) and inter-token latency (ITL), under highly dynamic and heterogeneous workloads~\cite{agrawal2024taming,kwon2023efficient}. As a result, designing efficient inference systems has become a critical challenge for modern AI infrastructure.

To overcome the computational and memory limitations of large models, modern LLM serving systems increasingly rely on distributed execution across multiple GPUs. Tensor Parallelism (TP), introduced by Megatron-LM~\cite{shoeybi2019megatron}, partitions transformer computations across devices by sharding model weights and intermediate activations, reducing the computation and memory requirements per GPU. Pipeline Parallelism (PP), explored in systems such as GPipe~\cite{huang2019gpipe} and PipeDream~\cite{harlap2018pipedream}, partitions model layers across devices and enables models that exceed the memory capacity of a single accelerator. More recently, hybrid parallelism (HB), combining TP and PP, has been adopted in large-scale deployments to balance computation, memory capacity, and communication overhead~\cite{narayanan2021efficient,aminabadi2022deepspeed}. These parallelism strategies form the foundation of modern LLM inference frameworks, including TensorRT-LLM~\cite{tensorrtllm_docs}, vLLM~\cite{vllm_docs}, and NVIDIA Dynamo~\cite{nvidia_dynamo}.

Despite extensive progress in distributed LLM serving, selecting an appropriate parallelism strategy remains challenging because performance depends on complex interactions among computation, communication, memory bandwidth, KV-cache capacity, and workload characteristics. Existing evaluations typically compare parallelism strategies through empirical measurements under fixed model configurations, batch sizes, and hardware settings. While such evaluations provide valuable insights, they often fail to explain the underlying mechanisms that determine performance scalability. In practice, the dominant bottleneck of LLM inference can shift significantly across workloads: prefill execution may be limited mainly by matrix computation, whereas decoding may become constrained by memory bandwidth, KV-cache accesses, and repeated synchronization overhead~\cite{kwon2023efficient,agrawal2024taming}. Consequently, conclusions obtained from a single workload configuration often do not generalize to different models, sequence lengths, or deployment environments.

This challenge is further amplified by the fundamentally different characteristics of the prefill and decoding phases. During prefill, the system processes the entire input sequence simultaneously and benefits from large matrix operations with high arithmetic intensity. In contrast, decoding generates tokens sequentially, requiring repeated execution of the model for every generated token and introducing frequent communication and synchronization operations. For example, TP reduces computation per GPU but introduces collective communication after transformer components, while PP reduces communication volume but suffers from pipeline bubbles when the amount of useful work is insufficient to keep all stages occupied. Although recent serving systems employ advanced scheduling techniques, including continuous batching, KV-cache management, and phase-aware execution~\cite{kwon2023efficient,agrawal2024taming,zhong2024distserve,patel2024splitwise}, most systems still rely on a fixed parallelism configuration across prefill and decode phases. Therefore, understanding how parallelism overhead evolves across different inference phases remains an open challenge.

In this work, we address this challenge by developing a unified analytical framework for modeling distributed LLM inference performance. Unlike prior studies that focus on empirical benchmarking or isolated optimization techniques, our framework captures the fundamental latency components introduced by TP, PP, and HB across both prefill and decode execution. Specifically, we model TP communication overhead using collective communication costs, characterize pipeline bubble effects for autoregressive inference, and integrate computation, communication, and pipeline overhead into a unified performance model parameterized by model architecture, sequence length, batch size, GPU configuration, and interconnect topology. Our goal is to provide a fundamental understanding of the scalability limits of distributed LLM inference and enable systematic reasoning about the trade-offs among different parallelism strategies.
\vspace{-3mm}
\subsection{Related Works}
To enable efficient execution of LLMs, prior research has extensively explored distributed parallelism strategies that partition computation and memory requirements across multiple accelerators. TP, introduced in Megatron-LM~\cite{shoeybi2019megatron}, distributes the computation of transformer layers by partitioning weight matrices across GPUs and synchronizing intermediate activations through collective communication operations. While TP effectively reduces per-device computation and enables high throughput within tightly coupled GPU systems, its scalability is limited by frequent communication overhead introduced by synchronization operations. PP, investigated in systems such as GPipe~\cite{huang2019gpipe} and PipeDream~\cite{harlap2018pipedream}, addresses memory scalability by partitioning model layers across sequential pipeline stages. However, PP introduces pipeline bubbles during pipeline fill and drain periods, reducing hardware utilization when the amount of available parallel work is insufficient. To overcome the limitations of TP and PP, HB strategies combining TP, PP, and Expert Parallelism (EP) have been proposed for large-scale model deployment~\cite{narayanan2021efficient,aminabadi2022deepspeed,fedus2022switch}. Although these implementations provide effective mechanisms for scaling model execution, they primarily focus on improving system scalability and implementation efficiency rather than developing a unified understanding of the performance trade-offs among different parallelism strategies.

Beyond designing parallel execution techniques, analytical performance modeling has been extensively studied in high-performance computing and distributed deep learning to understand scalability limitations. Classical communication models, including the Hockney $\alpha-\beta$ model~\cite{10.1016MPP} and the LogP model~\cite{647225}, characterize communication latency as a combination of fixed startup overhead and bandwidth-dependent transmission cost. These models have been widely adopted for analyzing collective communication operations such as AllReduce and AllGather, including implementations provided by libraries such as NVIDIA collective communications library (NCCL)~\cite{nvidia_nccl}. Similarly, the roofline model~\cite{10.1145/1498765.1498785} provides a framework for identifying whether applications are limited by computation throughput or memory bandwidth. Recent distributed deep learning systems have extended these ideas to analyze communication and computation overlap in large-scale training workloads. However, existing analytical models are primarily designed for traditional HPC applications or distributed training scenarios and do not capture the unique characteristics of autoregressive LLM inference, where TP communication occurs repeatedly during decoding, pipeline bubbles depend on inference workload characteristics, and performance varies significantly between prefill and decode phases.

From serving perspectives, numerous techniques have been introduced, including continuous batching, KV-cache management, memory optimization, request scheduling, and workload-aware execution. Frameworks such as vLLM~\cite{kwon2023efficient}, TensorRT-LLM~\cite{tensorrtllm_docs}, Sarathi-Serve~\cite{agrawal2024taming}, DistServe~\cite{zhong2024distserve}, Splitwise~\cite{patel2024splitwise}, Mooncake~\cite{qin2024mooncake}, and NVIDIA Dynamo~\cite{nvidia_dynamo} have demonstrated significant improvements in practical LLM deployment by optimizing resource utilization and reducing serving latency. In parallel, several studies have proposed analytical approaches to model inference latency based on GPU utilization, memory bandwidth, KV-cache behavior, and workload characteristics. However, existing efforts generally focus on either improving serving efficiency empirically or modeling individual performance components in isolation. To the best of our knowledge, there is no unified analytical framework that jointly captures TP communication overhead, PP bubble effects, and their hybrid execution across both prefill and decode phases. This limitation prevents a fundamental understanding of how parallelism choices affect LLM inference scalability across different models, workloads, and hardware configurations, which motivates the analytical framework developed in this work backed by experiment results.

\section{Parallelism Strategies} \label{sec:parallel}
Existing LLM serving frameworks primarily employ three parallelism strategies: TP, PP, and HB, where HB combines TP and PP. Each strategy exposes a fundamentally different trade-off between computation, communication, and hardware utilization. TP partitions the computation of individual transformer layers across multiple GPUs, which enables concurrent execution at the cost of frequent collective communication and synchronization. In contrast, PP partitions the model layers across GPUs. It effectively reduces the inter-GPU communication to point-to-point activation transfers while introducing pipeline bubbles due to pipeline fill and drain effects. HB seeks to balance these trade-offs by combining TP within each pipeline stage and PP across stages, thereby reducing the limitations of either approach alone. Consequently, the relative efficacy of these strategies depends on multiple factors, including batch size, sequence length, model architecture, and GPU topology. This motivates a detailed analysis of their computation, communication, and pipeline characteristics. The following sections describe each parallelism strategy and develop analytical models that characterize their latency behavior during both the prefill and decoding phases.
% \begin{figure}
%     \centering
%     \includesvg[width=\linewidth]{figs/TP_SVG}
%     \caption{Tensor Parallelism}
%     \label{fig:TP}
% \end{figure}

\begin{figure}
    \centering
    \includegraphics[width=\linewidth]{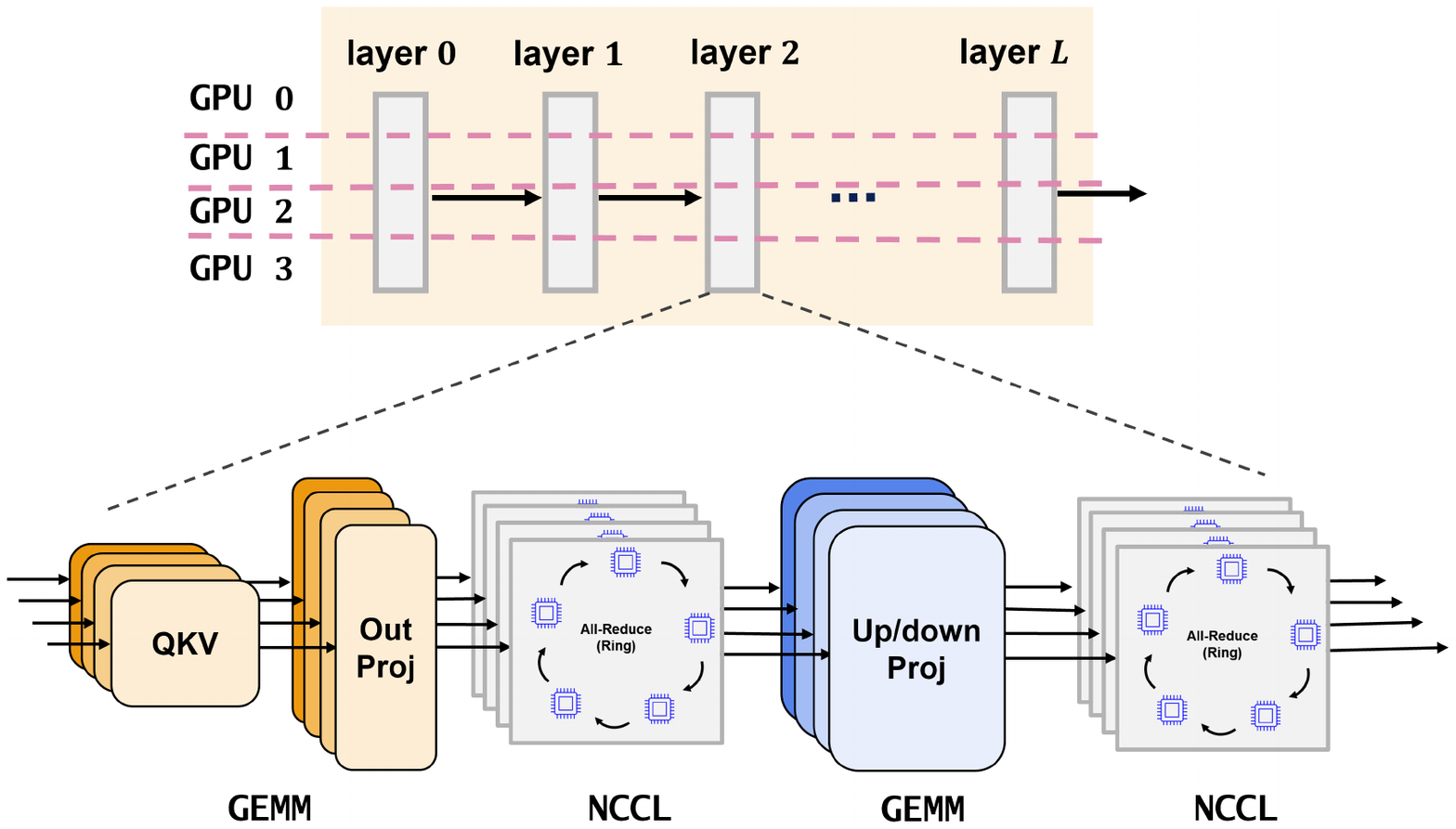}
    \vspace{-15mm}
    \caption{Tensor Parallelism Data Flow.}
    \vspace{-5mm}
    \label{fig:TP}
\end{figure}

\subsection{Tensor Parallelism}\label{sec:TP}
Tensor Parallelism is one of the most widely adopted techniques for scaling LLM inference across multiple GPUs. As shown in Fig. \ref{fig:TP}, TP partitions the weight matrices of each transformer layer across multiple GPUs, allowing the computation of large matrix multiplications to be distributed among participating devices. This partitioning reduces the amount of work performed by each GPU and enables larger models to be served efficiently. However, because each GPU computes only a portion of the layer output, intermediate results must be exchanged and synchronized through collective communication operations. Consequently, TP introduces a fundamental trade-off: increasing the TP degree reduces computation per GPU but increases communication between GPUs. To better understand this trade-off, let us define the total latency in TP as
\begin{equation}\label{TP_total_latency}
T^{\rm total}_{\rm TP} =T^{\rm comp}_{\rm TP} +T^{\rm comm}_{\rm TP},
\end{equation}
where $T^{\rm comp}_{\rm TP}$ and $T^{\rm comm}_{\rm TP}$ are the corresponding compute and communication terms, respectively. We further note that the compute latency for a transformer with $L$ layers is $T^{\rm comp}_{\rm TP}  = L T^{\rm (s), comp}_{\rm TP}$, where 
\begin{equation}\label{TP_layer_latency}
T^{\rm (s), comp}_{\rm TP} =\frac{1}{P_{\rm TP}}\frac{FLOPs_{\rm layer} (B,S)} {\gamma}
\end{equation}
is the compute latency in each layer, $FLOPs_{\rm layer} (B,S)$ is the total floating-point operations required in each layer for a batch size $B$ and sequence length $S$, $P_{\rm TP}$ is the TP degree, and $\gamma$ denotes the effective compute throughput of a single GPU (FLOPs/s). The expression in \eqref{TP_layer_latency} reflects the key property of TP, where partitioning model weights across more GPUs reduces the computation assigned to each GPU approximately in proportion to $P_{\rm TP}$.

\paragraph{Communication Latency Modeling:} 
NCCL \cite{nvidia_nccl} is the de facto communication backend in multi-GPU LLM serving. It provides highly optimized communication primitives over high-speed interconnects such as NVLink and InfiniBand. NCCL abstracts low-level communication procedures and automatically selects efficient communication algorithms based on the underlying hardware topology as well as the workload. NCCL organizes participating GPUs into a communicator, which defines the communication domain for collective operations. Before any communication can occur, a communicator is created using APIs such as \verb|ncclCommInitAll|  for single-process execution or \verb|ncclCommInitRank| for distributed multi-process environments. NCCL also supports grouped execution through \verb|ncclGroupStart| and  \verb|ncclGroupEnd|, allowing multiple communication operations to be launched together, thereby reducing kernel launch overhead and improving communication efficiency.

During inference, only a subset of NCCL primitives is utilized. In particular, AllReduce is the dominant communication primitive in Megatron-style TP \cite{shoeybi2019megatron}, where each GPU computes partial activations that must be aggregated across all participating GPUs after the attention and feed-forward blocks of every transformer layer. Other collectives, including \verb|ReduceScatter|, \verb|AllGather|, \verb|Broadcast|, and \verb|Reduce|, are employed in different distributed training and inference frameworks, while \verb|Send| and \verb|Recv| are primarily used in PP for point-to-point activation transfers between adjacent pipeline stages \cite{hu2025demystifying}. Because transformer decoding performs these collective operations repeatedly for every layer and every generated token, the efficiency of inter-GPU communication directly affects the scalability of TP for inference. Consequently, accurately modeling communication latency is essential for understanding the trade-off between reduced computation and increased synchronization overhead as TP degree increases.

In Megatron-style TP \cite{shoeybi2019megatron}, \verb|AllReduce| is a dominant communication operation, where partial activations computed on different GPUs must be synchronized after the attention and feed-forward blocks of every transformer layer. Although TP reduces the computational workload assigned to each GPU, it introduces these recurring synchronization events that can significantly impact inference performance. Specifically, each transformer layer performs two activation synchronization steps, and these collective operations are executed for every layer and every generated token during decoding. As the TP degree increases, the amount of computation executed by each GPU decreases; while the synchronization overhead increases, which makes communication an increasingly important component of the overall latency budget. Our profiling results in Section \ref{sec:exp_TP} show that, despite the reduced per-GPU message size at higher TP degrees, inter-GPU communication can account for $30–40\%$ of the end-to-end decoding latency. This behavior indicates that communication latency is governed not only by the volume of transmitted data but also by the frequency of collective operations and the latency characteristics of the underlying interconnect. To better understand this behavior and predict the scaling behavior of TP, we next develop an analytical model for inter-GPU communication latency that captures both message transmission costs and collective synchronization overhead. To this end, we define 
\begin{equation} \label{eq:activation}
|M|= BSd_{\rm hidden}b,
\end{equation}
the message size transferred between GPUs, $S=\rm ISL$ for prefill, and $S=1$ for decoding, $d_{\rm hidden}$ is the model's hidden dimension, and $b$ is the number of bytes for activations. We observe that the message size scales with batch size. Then, according to Megatron-style TP \cite{shoeybi2019megatron}, the inter-GPU communication latency is given by:
\begin{equation}\label{T_NCCL}
T^{\rm comm}_{\rm TP} = LT^{\rm (s), comm}_{\rm TP}
\end{equation}
where $L$ is the number of transformer's layers, 
\begin{equation}\label{T_NCCL1}
T^{\rm (s), comm}_{\rm TP}=k(P_{\rm TP}-1) \left[\alpha +\left(\frac{1}{P_{\rm TP}}\right)\left(\frac{|M|}{\beta}\right)\right],
\end{equation}
$k$ is the number of collectives that occur per layer, $\alpha$ is the network's intrinsic latency arising mainly from kernel launch and synchronization,  and $\beta$ is the effective inter-GPU NVLink bandwidth. This expression indicates that TP communication grows linearly with batch size and sequence length. Importantly, while increasing the TP degree reduces the communicated payload per GPU, it also increases the number of participating GPUs and the associated synchronization overhead. 

The expressions in \eqref{TP_layer_latency} and \eqref{T_NCCL1}, expose the fundamental compute-communication trade-off in TP. In particular, computation decreases by $P_{\rm TP}$, meanwhile, communication decreases only partially with message size while synchronization overhead increases with $P_{\rm TP}$. This competition ultimately yields the optimal TP degree, to balance compute and communication in order to maximize throughput.

\subsection{Pipeline Parallelism}
\begin{figure}
    \centering
    \includegraphics[width=.5\linewidth]{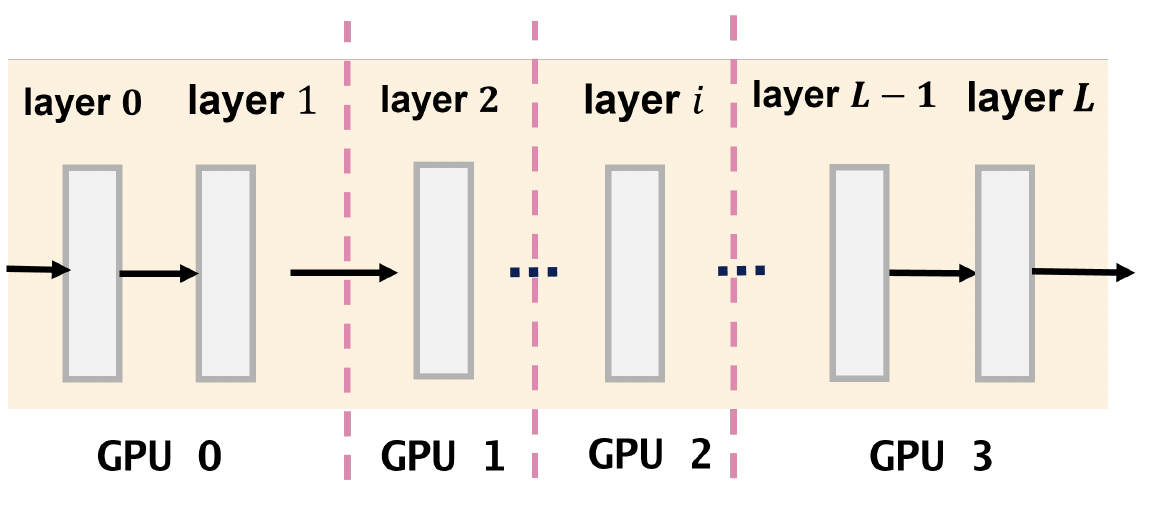}
    \caption{Pipeline Parallelism Data Flow.}
    \vspace{-5mm}
    \label{fig:PP}
\end{figure}

Pipeline Parallelism adopts a fundamentally different approach from TP by partitioning the transformer layers across multiple GPUs rather than splitting individual layer computations. As shown in Fig. \ref{fig:PP}, each GPU is responsible for an $L_{\rm PP}$ contiguous subset of layers, and activations are forwarded from one stage to the next as execution progresses through the model. This approach significantly reduces the collective communication overhead associated with TP synchronization, as GPUs primarily exchange activations only between neighboring pipeline stages. Furthermore, PP enables serving larger models by distributing model weights across multiple devices. However, these benefits come at the cost of a performance inefficiency referred to as a pipeline bubble. Since different pipeline stages must wait for activations to arrive before computation can begin, some stages may remain idle during pipeline startup and drain phases, resulting in underutilized hardware resources. The magnitude of this idle time depends on both the pipeline depth and the number of batches available to keep the pipeline occupied.
% \begin{figure}[ht!]
%     \centering
%     \includesvg[width=.5\linewidth]{figs/PP}
%     \caption{Pipeline Parallelism.}
%     \label{fig:PP}
% \end{figure}
To quantify this idle latency, let us define the total latency of PP as
\begin{equation}\label{PP_total_latency}
T^{\rm total}_{\rm PP} = T^{\rm comp}_{\rm PP} +T^{\rm comm}_{\rm PP} + T^{\rm bubble}_{\rm PP},
\end{equation}
where $T^{\rm comp}_{\rm PP}$, $T^{\rm comm}_{\rm PP}$, and $T^{\rm bubble}_{\rm PP}$ denote the compute, communication and bubble latency of PP, respectively. To quantify each component, we note that before all stages become active, the pipeline must be filled, which requires $P_{\rm PP}-1$ stage intervals. Similarly, after the final batch enters the pipeline, an additional $P_{\rm PP}-1$ stage intervals are required to drain the remaining work. Therefore, the total pipeline execution time can be modeled as
\begin{equation}\label{eq:pipe}
T^{\rm total}_{\rm PP}=(W+P_{\rm PP}-1)T^{(\rm s)}_{\rm PP},
\end{equation}
where $W=BS$ is the number of pipeline tasks (i.e., token-level work units) processed in one iteration ($S=\rm ISL$ for prefill and $S=1$ for decode), and 
\begin{equation}\label{eq:Ts_PP}
T^{(\rm s)}_{\rm PP} = T^{(\rm s), comp}_{\rm PP}+ T^{(\rm s), comm}_{\rm PP},
\end{equation}
denotes the execution time of a single pipeline stage, including both computation and communication. Assuming the model is evenly partitioned across $P_{\rm PP}$ pipeline stages, each stage executes $L/P_{\rm PP}$ transformer layers. Given $FLOPs_{\rm layer} (B,S)$  as floating-point operations required by one transformer layer for batch size $B$ and sequence length $S$, $\gamma$ the effective computational throughput of a single GPU, the compute latency of one pipeline stage is given by
\begin{equation}\label{PP_s_comp}
T^{\rm (s), comp}_{\rm PP} = \frac{L}{P_{\rm PP}}\frac{FLOPs_{\rm layer} (B,S)}{\gamma}.
\end{equation}
Then, PP's total compute latency, $T^{\rm comp}_{\rm PP}$, is given by
\begin{equation}\label{eq:pp_comp}
T^{\rm comp}_{\rm PP}=WT^{(\rm s), comp}_{\rm PP}.
\end{equation}
From \eqref{eq:pp_comp}, compute latency increases with both the batch size and sequence length through $FLOPs_{\rm layer} (B,S)$. Therefore, prefill ($S=\rm ISL$) incurs substantially higher computation than decoding ($S=1$), while increasing the pipeline degree reduces the computation assigned to each stage approximately in proportion to $1/P_{\rm PP}$.

To quantify the communication and bubble latency, we note that during inference, intermediate activations produced by one stage must be transmitted to the next stage before computation can proceed. Consequently, PP introduces inter-GPU point-to-point communication at every stage boundary. Unlike TP, PP does not shard activations across devices; therefore, $|M|$ scales linearly with both batch size and sequence length. Assuming a standard latency-bandwidth communication model, the communication latency between two adjacent stages can be expressed as
\begin{equation} \label{eq:PP_comm}
T^{(\rm s), comm}_{\rm PP}=\alpha+\frac{|M|}{\beta}.
\end{equation}
Note that $T^{(\rm s), comm}_{\rm PP}$ scales proportionally with batch size and sequence length through $|M|$. Consequently, prefill requests with long input sequences ($S=\rm ISL$) incur significantly higher communication costs than decode iterations ($S=1$). Note that for small batch size, the activation size is relatively small, causing the startup latency term $\alpha$ to dominate the communication cost. In this regime, increasing the batch size improves communication efficiency by amortizing the fixed startup overhead across a larger activation payload. As the batch size continues to increase, the bandwidth term becomes dominant and communication latency grows approximately linearly with message size.

In addition to inter-GPU communication latency, PP suffers from pipeline bubbles arising from imperfect pipeline occupancy. In order to quantify $T^{\rm bubble}_{\rm PP}$, from \eqref{eq:pipe}, we note that only $W T^{(\rm s)}_{\rm PP}$ is the useful interval while the rest of $(P_{\rm PP} -1)T^{(\rm s)}_{\rm PP}$ is related to idle time (i.e. pipeline bubble). Therefore, the communication latency and pipeline bubble latency are given by
\begin{eqnarray}
    T^{\rm comm}_{\rm PP}&=&WT^{(\rm s), comm}_{\rm PP}, \label{eq:comm}\\
T^{\rm bubble}_{\rm PP}&=&(P_{\rm PP}-1)(T^{(\rm s), comp}_{PP}+ T^{(\rm s), comm}_{\rm PP}), \label{eq:bubble1}
\end{eqnarray}
% \begin{equation}\label{eq:comm}
% T^{\rm comm}_{\rm PP}=WT^{(\rm s), comm}_{\rm PP},
% \end{equation}
% \begin{equation}\label{eq:bubble1}
% T^{\rm bubble}_{\rm PP}=(P_{\rm PP}-1)(T^{(\rm s), comp}_{PP}+ T^{(\rm s), comm}_{\rm PP}),
% \end{equation}
and therefore, the bubble fraction is given by 
\begin{equation}\label{eq:bubble_frac}
\rho^{\rm bubble}_{\rm PP} = \frac{T^{\rm bubble}_{\rm PP}}{T^{\rm total}_{\rm PP}}=\frac{P_{\rm PP}-1}{W+P_{\rm PP}-1}.
\end{equation}
From \eqref{eq:bubble1} and  \eqref{eq:bubble_frac} we observe that $T^{\rm bubble}_{\rm PP}$ increases with the PP degree, however $\rho^{\rm bubble}_{\rm PP}$ decreases with batch size. Intuitively, the fill/drain overhead remains fixed at $P_{\rm PP}-1$ stage intervals, while increasing the number of batches allows this fixed cost to be amortized over a larger amount of useful work.

\subsection{Hybrid Parallelism}
In this section, we consider an HB configuration consisting of $P_{\rm PP}$ pipeline stages, each hosting $L_{\rm PP}$ consecutive layers of the model. The computational load of these layers is distributed among $P_{\rm TP}$ GPUs within each stage as described in Section \ref{sec:TP}. Therefore, the total number of GPUs used for parallelism is given by $P=P_{\rm PP}P_{\rm TP}$. For both prefill and decode, the total latency is given by 
\begin{eqnarray}
    T^{\rm total}_{\rm HB} &=& T^{\rm comp}_{\rm HB} +T^{\rm comm}_{\rm PP} + T^{\rm bubble}_{\rm PP}, \nonumber \\
&=& T^{\rm comp}_{\rm TP} +L_{\rm PP}T^{\rm (s), comm}_{\rm TP} +T^{\rm comm}_{\rm PP} + T^{\rm bubble}_{\rm PP},
\end{eqnarray}\label{HB_total_latency}
where $T^{\rm comp}_{\rm TP}$ is the aggregated compute latency carried out by TP (see \eqref{T_NCCL}), $T^{\rm comm}_{\rm PP}$ denotes the communication latency between $P_{\rm PP}$ pipeline stages (see \eqref{eq:comm}), and $T^{\rm bubble}_{\rm PP}$ is the bubble latency caused by inefficiencies in pipeline parallelism, which is described in more detail below.
\paragraph{Hybrid Parallelism Communication Modeling:} During inference in HB, communication arises from two distinct sources. First, GPUs within the same TP group exchange partial activations through collective communication operations such as \verb|AllReduce|, \verb|AllGather|, and \verb|ReduceScatter|. Second, activations are forwarded between adjacent PP stages through point-to-point communication. The per stage latency is defined as 
\begin{equation}\label{eq:HB_stage}
T^{(\rm s)}_{\rm HB}= T^{\rm (s), comp}_{\rm HB} + T^{\rm (s), comm}_{\rm PP}.
\end{equation}

Similar to PP, the total pipeline execution time is given by
\begin{equation}\label{eq:HB}
T^{\rm total}_{\rm HB}=(W+P_{\rm PP}-1)T^{(\rm s)}_{\rm HB},
\end{equation}
and the bubble latency as
\begin{equation}\label{eq:bubble}
T^{\rm bubble}_{\rm HB}=(P_{\rm PP}-1)T^{(\rm s)}_{\rm HB}.
\end{equation}
Therefore, the bubble fraction is given by 
\begin{equation}\label{eq:bubble_frac_HB}
\rho^{\rm bubble}_{\rm HB} = \frac{T^{\rm bubble}_{\rm HB}}{T^{\rm total}_{\rm HB}}=\frac{P_{\rm PP}-1}{W+P_{\rm PP}-1}.
\end{equation}
It is worth noting that \eqref{eq:bubble_frac_HB} and \eqref{eq:bubble_frac} are identical. They are only a function of $W$, pipeline depth,  and independent of TP degrees. In particular, TP changes $T^{(\rm s)}_{\rm HB}$ but not bubble latency. Furthermore, unlike TP, which only modifies the execution time of an individual pipeline stage through distributed computation and collective communication, the pipeline fill/drain process depends solely on the $P_{\rm PP}$ and the amount of work entering the pipeline. Consequently, introducing TP changes the absolute bubble latency through the stage execution time $T^{(\rm s)}_{\rm HB}$, but does not alter $\rho^{\rm bubble}_{\rm HB}$ in \eqref{eq:bubble_frac_HB}. This implies that TP can reduce the duration of each pipeline interval, whereas improved pipeline utilization can be achieved via only increasing the amount of work $W$ or reducing the pipeline depth $P_{\rm PP}$.
\begin{table}[h]
\centering
\caption{Sample Kernels.}
\label{tab:kernels}
\begin{tabular}{l||l}
\hline
\textbf{Category} & \textbf{Kernel Name} \\
% \midrule
\hline
\multirow{8}{*}{\texttt{GEMM}} & 
\verb|ampere_bf16_s16816gemm_bf16_256x128_ldg8_relu_f2f_stages_64x3_tn|\\
& \verb|ampere_bf16_s16816gemm_bf16_64x64_ldg8_relu_f2f_stages_64x5_tn| \\
& \verb|ampere_bf16_s16816gemm_bf16_256x128_ldg8_f2f_stages_64x3_tn| \\
& \verb|ampere_bf16_s16816gemm_bf16_128x64_ldg8_f2f_stages_64x4_tn| \\
& \verb|ampere_bf16_s16816gemm_bf16_64x64_sliced1x2_ldg8_f2f_stages_64x6_tn| \\
 & \verb|ampere_bf16_s16816gemm_bf16_64x64_sliced1x2_ldg8_relu_f2f_stages_64x5_tn| \\
 & \verb|ampere_bf16_s16816gemm_bf16_64x64_sliced1x2_ldg8_f2f_stages_64x5_tn| \\
\hline
\multirow{4}{*}{\texttt{NCCL}} &\verb|ncclDevKernel_AllReduce_Sum_bf16_RING_LL| \\
 &\verb|ncclDevKernel_AllGather_RING_LL| \\
 &\verb|ncclDevKernel_SendRecv| \\
 &\verb|void vllm::cross_device_reduce_1stage|\\
\hline
\multirow{4}{*}{\texttt{Attention}}
&\verb|triton_poi_fused_cat_3|\\
&\verb|triton_poi_fused_cat_4|\\
&\verb|vllm::reshape_and_cache_flash_kernel|\\
&\verb|flash::flash_fwd_splitkv_kernel|\\
&\verb|flash::flash_fwd_splitkv_combine_kernel|\\

\hline
\multirow{6}{*}{\texttt{Other}}
&\verb|triton_red_fused__to_copy_add_mean_mul_pow_rsqrt_2|\\
 &\verb|triton_red_fused__to_copy_add_mean_mul_pow_rsqrt_0|\\
 &\verb|triton_poi_fused_mul_silu_1|\\
 &\verb|at::native::unrolled_elementwise_kernel|\\
 &\verb|at::native::index_elementwise_kernel|\\
 &\verb|Memcpy DtoH|, \verb|Memcpy HtoD|, \verb|Memcpy DtoD|\\
\hline
% \bottomrule
\end{tabular}
\end{table}

\section{Experiments}
In this section, we provide experiments to evaluate modern LLMs deployed across multiple GPUs. The primary objective of our evaluation is to characterize the performance trade-offs among TP, PP, and HB during both prefill and decoding phases of autoregressive inference. We first describe the experimental setup, including the hardware platform, models, serving framework, and evaluation methodology. We then analyze TP, PP, and HB independently by decomposing execution time into four major kernels: matrix multiplication (\verb|GEMM|) operations, attention computation (\verb|Attention|), communication overhead (\verb|NCCL|), and miscellaneous operations (\verb|Other|) such as kernel launches, memory management, and framework overhead. Samples of such kernels are provided in Table \ref{tab:kernels}. For each parallelism strategy, we evaluate both TTFT and ITL across different batch sizes. The results in this section provide insights on how computation, communication, and pipeline utilization collectively determine the scalability and efficiency of distributed LLM inference.

\subsection{Experiment Setup}
Throughout the experiments, we employ Qwen2.5-32B-Instruct and Qwen2.5-14B-Instruct \cite{qwen2025qwen25technicalreport} on $8\times$ NVIDIA A100 SXM4 (80 GB) \cite{NvdaA100}. The GPUs are connected through NVLink, providing an aggregate bidirectional bandwidth of 600 GB/s between neighboring GPUs, to enable high-throughput collective communication required by TP and PP for inference. Unless otherwise stated, inference is performed using FP16 precision with TP and PP degrees of {1,2,4,8}. All experiments are performed using vLLM v0.15.0 as an inference engine \cite{vllm15}.

We evaluate the efficiency of inference for both prefill and decoding across different batch sizes, while fixing the ISL and OSL. To characterize both prefill and decode performance, we report TTFT and ITL together with kernel-level execution breakdowns obtained through NVIDIA Nsight Systems \cite{Nsight_sys_v2026_5}, allowing us to separately quantify \verb|GEMM|, \verb|Attention|, \verb|NCCL|, and remaining runtime overhead. Detailed experiment parameters are provided in Table \ref{tab:setup}.

\begin{table}[h]
\centering
\caption{Experimental Parameters.}
\label{tab:setup}
\begin{tabular}{l||l}
\hline
\textbf{Parameter} & \textbf{Value} \\
\hline
GPU Platform & 8 $\times$ NVIDIA A100 SXM4 (80 GB) \\
GPU Memory & 80 GB HBM2e per GPU \\
HBM Bandwidth & 2.0 TB/s per GPU \\
Peak Compute & 312 TFLOPS (FP16/BF16 Tensor Core) \\
Interconnect & NVLink 600 GB/s \\
Inference Framework & vllm v0.15.1 \\
Model & Qwen2.5-32B-Instruct \\
& Qwen2.5-14B-Instruct \\
Precision & FP16 \\
Parallelism Degrees & TP/PP $\in \{1,2,4,8\}$ \\
Batch Size & $B\in\{1,2,4,\ldots,128\}$ \\
$\rm ISL$ & 1000 tokens \\
$\rm OSL$ & 200 tokens \\
Profiler & NVIDIA Nsight Systems v2026.5.1 \\
\hline
\end{tabular}
\end{table}

\subsection{Tensor Parallelism Scaling Characteristics}\label{sec:exp_TP}

\begin{figure}[t!]
    \centering
    \psfrag{B}{$BB22$}
    \includegraphics[width=0.5\linewidth]{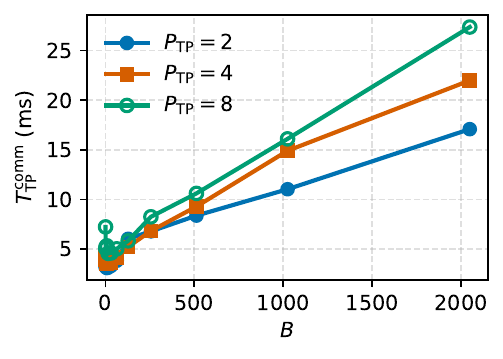}
    \caption{$T^{\rm comm}_{\rm TP}$ per token vs. batch size for different TP degrees. TP communication is dominated by startup latency at small batch sizes and gradually transitions to bandwidth-limited behavior at large batch sizes.}
    \vspace{-5mm}
    \label{fig:NCCL_plots}
\end{figure}

\paragraph{Communication Characterization:} In Fig. \ref{fig:NCCL_plots}, we plot $T^{\rm comm}_{\rm TP}$ per token vs batch size for various TP degrees. For fixed TP, NCCL has three competing operating regions that govern the behavior of TP. At small batch sizes, and according to \eqref{T_NCCL1}, the fixed communication latency term $\alpha$ dominates the overall communication cost. In this regime, the amount of data being exchanged is relatively small, and most of the communication time is spent initiating and coordinating collective operations rather than transferring payload data. As a result, increasing the batch size has only a limited impact on communication latency, since the fixed startup overhead is already the primary contributor. The system therefore operates in a latency-bound regime where communication cost is largely insensitive to message size.

As the batch size increases, the contribution of data transfer begins to grow and communication enters a transition region. In this regime, both $\alpha$ and the bandwidth-dependent transfer cost contribute meaningfully to $T^{\rm comm}_{\rm TP}$. Consequently, communication latency increases with batch size, but not yet in a strictly linear manner. The system is gradually shifting from being dominated by synchronization overhead toward being constrained by data movement. This region is particularly important because it often corresponds to the operating range of production serving systems, where both latency and throughput considerations must be balanced.

\begin{figure}[t!]
    \centering
    
    \begin{subfigure}[t]{0.49\textwidth}
        \centering
        \includegraphics[width=\linewidth]{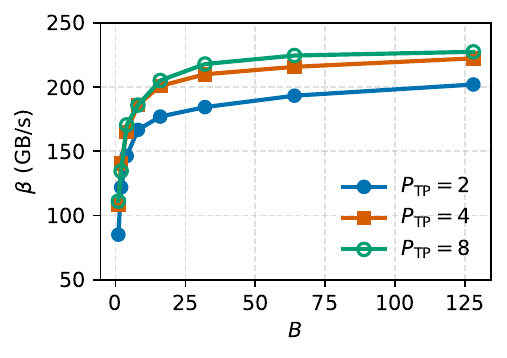}
        \vspace{-3mm}
        \caption{Prefill.}
        \label{fig:BW_prefill}
    \end{subfigure}
    \hfill
    \begin{subfigure}[t]{0.49\textwidth}
        \centering
        \includegraphics[width=\linewidth]{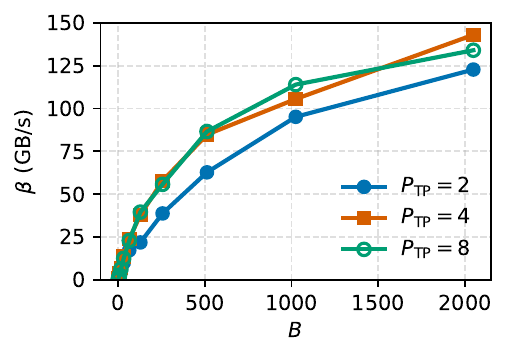}
        \vspace{-3mm}
        \caption{Decode.}
        \label{fig:BW_decode}
    \end{subfigure}

    \caption{Effective bandwidth vs. batch size for different TP degrees during (a) Prefill and (b) Decode. Bandwidth utilization grows with communication volume and eventually plateaus, revealing the transition from latency-dominated communication to sustainable inter-connect throughput. Decode remains more latency-limited than prefill due to smaller message sizes, resulting in lower effective bandwidth utilization.}
    \vspace{-5mm}
    \label{fig:HB_kernel_contribution}
\end{figure}

At sufficiently large batch sizes, $\alpha$ (i.e., startup latency) becomes negligible compared to the time required to move data across the interconnect. The communication cost is then dominated by the bandwidth term, causing latency to scale approximately linearly with message size. In this bandwidth-bound regime, further increases in batch size directly translate into higher communication overhead, and the achievable performance becomes limited by the effective inter-GPU bandwidth rather than synchronization latency. This observation highlights a key characteristic of TP, where small-batch workloads are primarily constrained by collective operation latency, large-batch workloads are fundamentally constrained by interconnect bandwidth, which makes efficient utilization of communication resources critical for sustaining scalability.
\paragraph{Effective Bandwidth:}
Figure \ref{fig:BW_prefill} illustrates the achieved interconnect bandwidth during the prefill vs batch size for different TP degrees. Although the A100 NVLink interconnect provides a theoretical peak bandwidth of approximately $600$ GB/s, the measured bandwidth never approaches this limit and instead saturates at roughly $250$ GB/s. This behavior reflects the fact that collective communication performance is constrained by more than the raw link bandwidth. In particular, AllReduce operations require synchronization among all participating GPUs, kernel launch overhead, protocol processing, and reduction operations before data transmission can fully utilize the interconnect. During prefill, the communicated activation tensors are relatively large (i.e., $S={\rm ISL}$), allowing bandwidth utilization to increase rapidly with batch size. However, once the communication pipeline becomes saturated, additional increases in message size provide little improvement in effective bandwidth, resulting in the observed plateau. Interestingly, higher TP degrees consistently achieve higher bandwidth utilization. To explain this, we note that although the case where $P_{\rm TP} = 8$ communicates fewer bytes per GPU due to tensor partitioning, the collective involves more participating GPUs and therefore generates a larger aggregate communication volume across the NVLink fabric. The communication subsystem consequently remains active longer to accommodate the communication among more GPU participants, amortizing startup costs and enabling the interconnect to operate closer to its sustainable throughput than the cases of $P_{\rm TP} = 2$ and $P_{\rm TP} = 4$.

During decoding, as shown in Fig.\ref{fig:BW_decode},  the achieved bandwidth is substantially lower, saturating at only approximately 150 GB/s, or about $25\%$ of the theoretical NVLink bandwidth. This reduction arises from the autoregressive nature of decoding, where only a single token is processed per iteration ($S=1$). Consequently, each of AllReduce operation transfers significantly smaller messages, causing communication to remain dominated by startup latency and synchronization overhead rather than bulk data movement. The smaller message sizes also delay the transition into the bandwidth-bound regime, explaining why the bandwidth plateau occurs at much larger batch sizes than during prefill. As in the prefill phase, the case where $P_{\rm TP} = 8$ consistently achieves higher bandwidth utilization than $P_{\rm TP} = 4$ and $P_{\rm TP} = 2$ because the increased number of GPUs participating in each collective produces greater aggregate network traffic and keeps the communication links busy for longer periods. Nevertheless, even the case $P_{\rm TP} = 8$ remains far from saturating the available NVLink bandwidth, indicating that, in TP, decoding is fundamentally latency-limited rather than bandwidth-limited. This observation explains why increasing the TP degrees often increases the communication latency observed in Fig.~\ref{fig:NCCL_plots}, where adding more GPUs introduces additional synchronization costs without generating sufficiently large messages to fully exploit the available interconnect bandwidth.

\begin{figure}[t!]
    \centering
    
\begin{subfigure}{\textwidth}
        \centering
        \includegraphics[width=\linewidth]{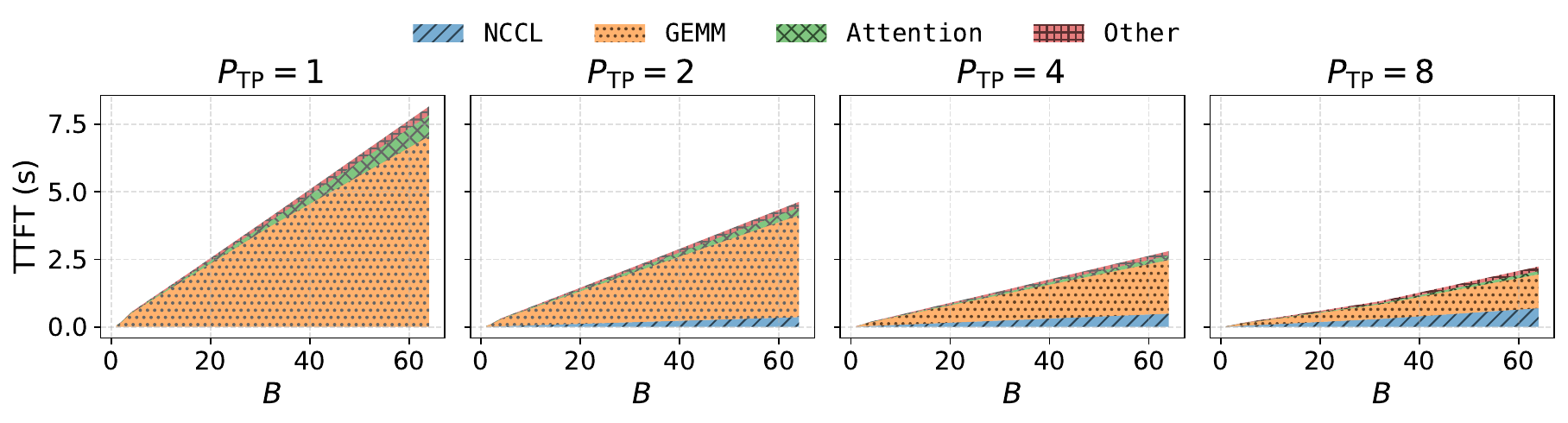}
        \caption{TTFT vs. Batch Size.}    
        \label{fig:TP_TTFT}
    \end{subfigure}
    
    % \vspace{0.5cm}
    \begin{subfigure}{\textwidth}
        \centering
        \includegraphics[width=\linewidth]{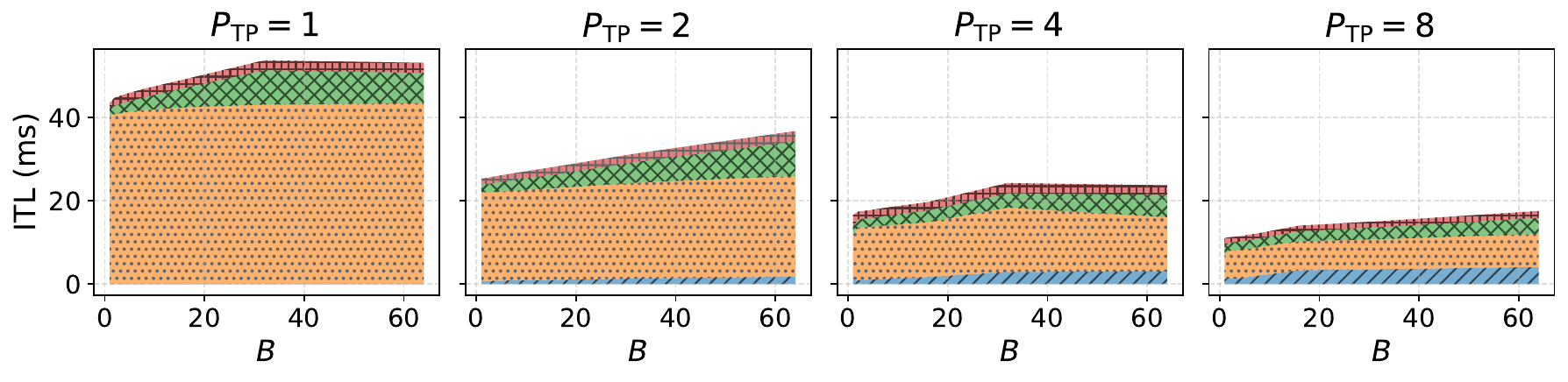}
        \caption{ITL vs. Batch Size.}
        \label{fig:TP_ITL}
    \end{subfigure}

    \caption{ Kernel-level breakdown of (a) TTFT and (b) ITL across different TP degrees for Qwen2.5 - 32B - Instruct. There is a fundamental compute-communication trade-off in TP, where computational savings from distributed  \texttt{GEMM} are partially offset by increased  \texttt{NCCL} synchronization overhead.}
    \vspace{-5mm}
\end{figure}

\paragraph{Compute-Communication Trade-off:} To evaluate the compute-communication trade-off, Fig. \ref{fig:TP_TTFT} shows the TTFT breakdown for different TP degrees. As TP degree increases from $1$ to $8$, the overall TTFT decreases because the computation performed by each GPU is reduced approximately proportionally with $P_{\rm TP}$. During prefill, every input token participates in the forward pass through all transformer layers, making matrix multiplications kernels (\verb|GEMMs|) the dominant contributor to latency. Partitioning matrix operations across more GPUs substantially reduces the computation assigned to each device, thereby resulting in a significant reduction in the \verb|GEMM| component and consequently the overall TTFT. This trend is consistent with \eqref{TP_layer_latency}, where $T^{\rm (s), comp}$ scales inversely with $P_{\rm TP}$.

For fixed $P_{\rm TP}$, TTFT increases almost linearly with batch size because the amount of computation grows proportionally with the number of processed tokens. During prefill, where $S=\rm ISL$, both \verb|GEMM| and \verb|NCCL| increase with the total number of input tokens (i.e., $BS$). Consequently, both $T^{\rm (s), comp}$ and $T^{\rm (s), comm}_{\rm TP}$ increase with batch size. However, the \verb|GEMM| component grows much faster than the \verb|NCCL| component since prefill is primarily compute-bound. Although communication increases with the activation size, the computational complexity of \verb|GEMM| scales much more rapidly. This makes \verb|GEMM| the dominant contributor to TTFT over the evaluated batch sizes.

Fig. \ref{fig:TP_ITL} presents the breakdown of ITL during decoding. Similar to TTFT, increasing $P_{\rm TP}$ reduces the overall ITL because each GPU performs only a fraction of the matrix multiplications. However, unlike prefill, the increase in ITL with batch size is much slower. During decoding, only a single token is generated per iteration ($S=1$), meaning that both the computational workload and the communicated activation size grow only with the batch size. Consequently, each decoding iteration performs significantly less computation than prefill, leading to a much slower increase in both the \verb|GEMM| and \verb|NCCL| components.
The communication overhead also increases more gradually because the activation tensors exchanged by \verb|NCCL| are substantially smaller during decoding. According to \eqref{eq:activation}, $|M|$ scales differently for both prefill and decode. As a result, the startup latency term $\alpha$ in \eqref{T_NCCL1} contributes a larger fraction of the communication time, causing \verb|NCCL| latency to increase more slowly with batch size than during prefill.

It is worth noting that increasing $P_{\rm TP}$ has opposite effects on computation and communication. On one hand, partitioning the model weights across more GPUs reduces the amount of computation assigned to each GPU, causing the \verb|GEMM| latency to decrease approximately proportionally with $P_{\rm TP}$. On the other hand, TP requires synchronization after every attention and feed-forward block. Although the communicated payload per GPU decreases by $P_{\rm TP}$, the number of participating GPUs and collective synchronization events increases with $P_{\rm TP}$. Consequently, the startup latency and synchronization overhead become increasingly dominant. This causes the \verb|NCCL| component to occupy a larger fraction of the total execution time. This behavior directly reflects the trade-off captured by the analytical model, where computation scales favorably with the $P_{\rm TP}$, whereas communication exhibits diminishing returns due to the increasing cost of collective synchronization.

\begin{figure}[t!]
    \centering
    
    \begin{subfigure}{0.45\textwidth}
        \centering
        \includegraphics[width=\linewidth]{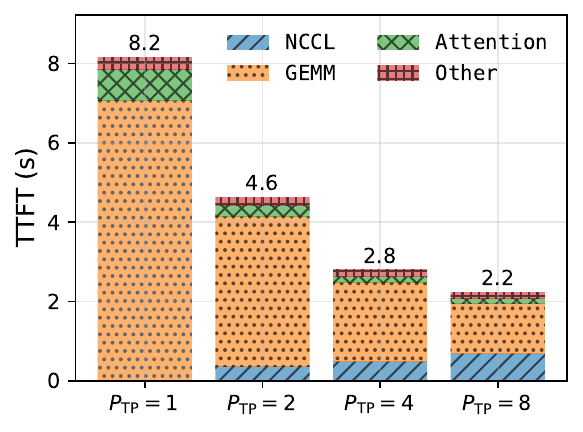}
        \caption{TTFT (s) vs. $P_{\rm TP}$}    
        \label{fig:ttft_actual}
    \end{subfigure}
    \hspace{-0.1cm}
    \begin{subfigure}{0.45\textwidth}
        \centering
        \includegraphics[width=\linewidth]{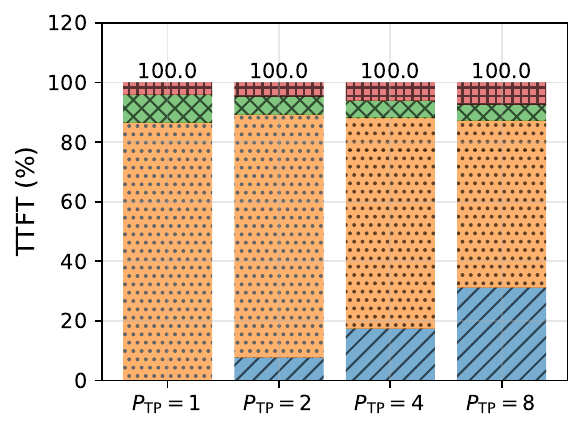}
        \caption{TTFT ($\%$) vs. $P_{\rm TP}$}
        \label{fig:ttft_percent}
    \end{subfigure}
    \vspace{0.5cm}
    \begin{subfigure}{0.45\textwidth}
        \centering
        \includegraphics[width=\linewidth]{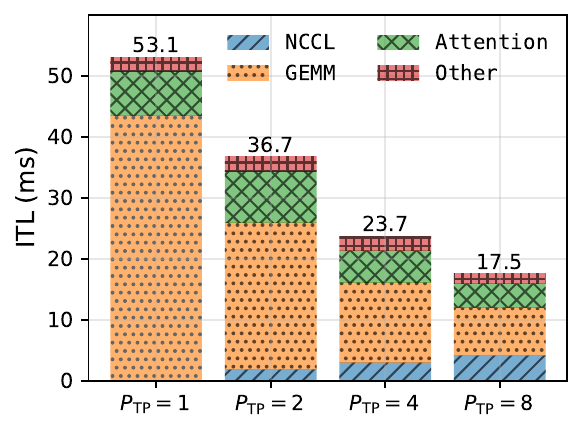}
        \caption{ITL (ms) vs. $P_{\rm TP}$}
        \label{fig:itl_actual}
    \end{subfigure}
    \hspace{-0.1cm}
    \begin{subfigure}{0.45\textwidth}
        \centering
        \includegraphics[width=\linewidth]{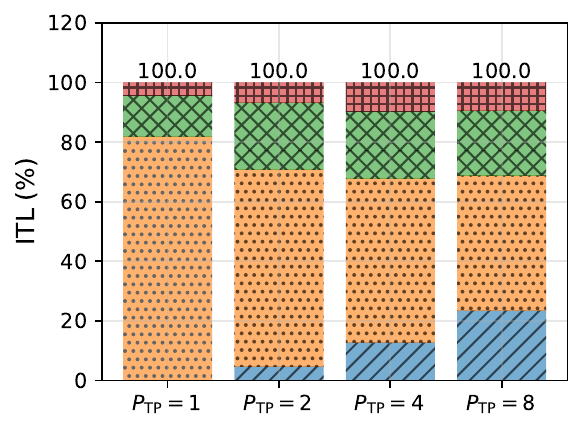}
        \caption{ITL ($\%$) vs. $P_{\rm TP}$}
        \label{fig:itl_percent}
    \end{subfigure}
\vspace{-5mm}
    \caption{Actual and normalized kernel contributions to TTFT and ITL across various TP degrees for Qwen2.5-32B-Instruct and $B=64$. Increasing TP degrees reduces both TTFT and ITL by lowering per-GPU computation, while  \texttt{NCCL} synchronization consumes an increasing fraction of latency, illustrating the trade-off between compute reduction and communication overhead.}
    \vspace{-5mm}
\end{figure}

To understand this trade-off, in Fig.~\ref{fig:ttft_actual}, as we increase TP degree, TTFT decreases. the reason is that the time spent in \verb|GEMM| kernels decreases substantially due to reduced per-GPU workload causing \verb|GEMM| operations to decrease approximately in proportion to the degree of parallelism. At the same time, \verb|NCCL| overhead grows steadily because more GPUs must participate in synchronization operations. Despite this increase in communication latency, the reduction in compute latency is initially larger than the communication penalty, resulting in an overall decrease in TTFT. In Fig.~\ref{fig:ttft_percent}, we show the contribution of each operation in TTFT. We observe that the total contribution of \verb|NCCL| and \verb|GEMM| remains the same across all TP degrees. However, as we increase TP degree, the communication overhead is traded for less \verb|GEMM| operation per GPUs, thereby resulting a significant reduction in TTFT. Note that at higher TPs, the communication overhead consumes almost $40\%$ of TTFT. A similar behavior is observed in ITL during decoding process as in Fig. ~\ref{fig:itl_actual}  and ~\ref{fig:itl_percent}. We observe that TP successfully accelerates both prefill and  decoding because the computational savings outweigh the additional communication overhead within the evaluated TP range.

In all the experiments in this section, we observed that, TP communication overhead originates from synchronization operations performed after distributed matrix multiplications. During these operations, each GPU exchanges partial results with all other participating GPUs through collective communication primitives. The latency of these communications depends on both the communication bandwidth available between GPUs and the number of synchronization events required throughout the model. As we increase TP degree, the frequency and cost of these communications become increasingly significant. Therefore, while \verb|GEMM| kernels become faster due to workload partitioning, \verb|NCCL| kernels consume a larger fraction of the total execution time. This behavior reflects a fundamental property of TP where reducing computation often requires additional coordination, and the benefits of parallel execution are ultimately constrained by communication efficiency.

{The result in this section indicates that increasing TP does not eliminate bottlenecks; rather, it shifts them. At low TP degrees, inference execution is dominated by computation, particularly \verb|GEMM| and \verb|Attention| kernels, making the system primarily compute-bound. As additional GPUs are introduced, the computational burden is reduced and communication becomes increasingly prominent. Eventually, a growing fraction of execution time is spent waiting for synchronization rather than performing useful computation. This transition represents a bottleneck shift from computation-bound execution toward communication-bound execution. Understanding this bottleneck transition is essential for selecting an appropriate TP configuration, as the optimal TP degree depends not only on model size and available hardware, but also on the balance between computation and communication within the target workload.}

\subsection{Pipeline Utilization and Bubble Overhead}
In PP, the latency is beyond the compute and communication inter-play. The pipeline inefficiencies introduces idle times (i.e., bubbles) that are particularly affect the latency during decoding. Since each active request contributes only a single token per decode iteration, the number of concurrently available work units is approximately equal to the active decode batch size. Consequently, for small decode batches $B \ll P_{\rm PP}$, the pipeline remains underutilized and a substantial fraction of execution time is spent in fill/drain phases. As the decode batch size increases, the number of active batches also increases, allowing all pipeline stages to be occupied. In the limit of large batch sizes $B \gg P_{\rm PP}$, the bubble fraction approaches zero.
In contrast, bubble overhead is generally less significant during prefill. Although the number of batches may remain unchanged, prefill workloads involve substantially longer input sequences ($S=\rm ISL$), resulting in much larger computational workloads per batch. Consequently, the fixed fill/drain overhead constitutes only a small fraction of the overall execution time, which makes pipeline bubbles largely negligible for long-context prefill workloads.

\begin{figure}[t!]
    \centering
    
    \begin{subfigure}{\textwidth}
        \centering
        \includegraphics[width=\linewidth]{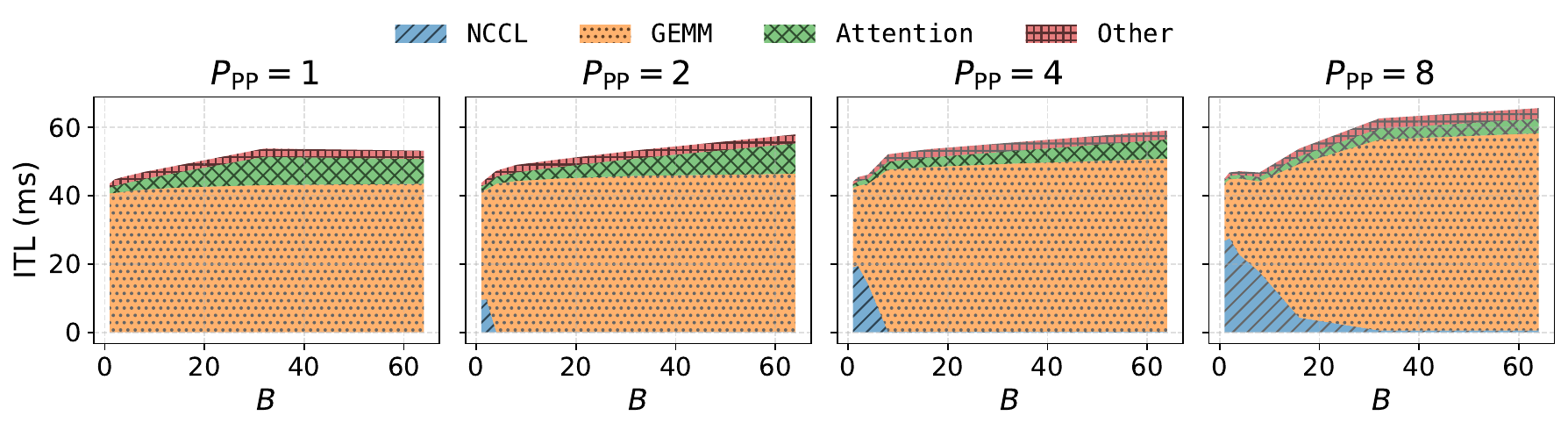}
        \caption{ITL (ms) vs. Batch Size.}    
        \label{fig:PP_ITL}
    \end{subfigure}
    
    % \vspace{0.5cm}
    \begin{subfigure}{\textwidth}
        \centering
        \includegraphics[width=\linewidth]{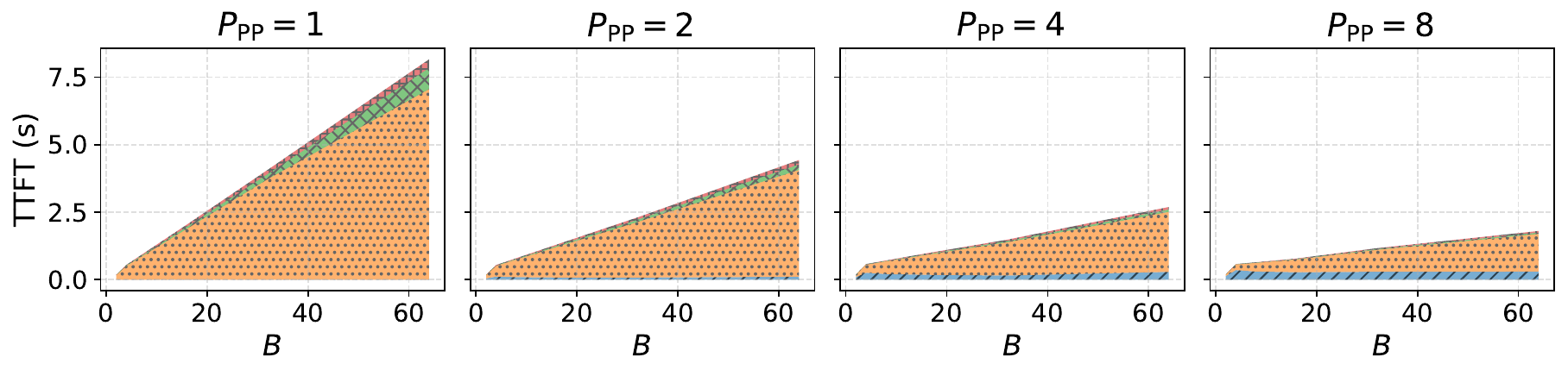}
        \caption{TTFT (s) vs. Batch Size.}
        \label{fig:PP_TTFT}
    \end{subfigure}

    \caption{Kernel-level decomposition of (a) ITL and (b) TTFT across different PP degrees for Qwen2.5 - 32B - Instruct. Decode performance is strongly influenced by pipeline bubbles at small batch sizes, whereas prefill workloads maintain high pipeline utilization and are primarily dominated by computation.}
    \vspace{-5mm}
\end{figure}

To understand the impact of pipeline bubbles during decoding, in Fig \ref{fig:PP_ITL}, we analyze the ITL across different pipeline degrees and batch sizes. Our results show that for $P_{\rm PP} > 1$, a significant fraction of the latency at small batch sizes is attributed to NCCL-related overhead. Unlike TP, these communication costs are not primarily caused by activation synchronization but rather reflect idle periods associated with pipeline bubbles. The effect becomes increasingly pronounced as the pipeline depth grows, since deeper pipelines require more stages to be filled before reaching steady-state execution. Consequently, the case $P_{\rm PP} = 8$ exhibits substantially larger bubble overhead than the case $P_{\rm PP} = 2$ or $P_{\rm PP} = 4$ at low batch sizes. As batch size increases, the number of simultaneously active decoding requests grows, generating more in-flight tokens and therefore more batches that can be distributed across the pipeline stages. These additional batches progressively fill the pipeline, reducing idle time and amortizing the bubble cost over a larger amount of useful work. Therefore, the NCCL-dominated bubble overhead gradually diminishes.

The behavior during prefill is markedly different and can be observed through the TTFT breakdown in Fig \ref{fig:PP_TTFT}. Unlike decoding, where only a limited number of tokens are active at any given time, prefill processes the entire input sequence simultaneously. Long input sequences naturally generate a large number of micro-batches and activation waves that continuously occupy the pipeline stages throughout execution. Consequently, the pipeline remains close to full utilization for most of the prefill phase, leaving little opportunity for pipeline bubbles to emerge. This behavior is reflected in Fig \ref{fig:PP_TTFT}, where the \verb|NCCL| component associated with bubble overhead is either significantly reduced or nearly absent across PP configurations. Instead, TTFT is primarily dominated by useful computation, including \verb|GEMM| and attention kernels. These observations highlight a fundamental distinction between decoding and prefill workloads: while decoding efficiency is highly sensitive to pipeline utilization and the availability of in-flight tokens, prefill inherently provides sufficient parallel work to keep deep pipelines occupied, which makes PP particularly well suited for prefill-dominated and long-context inference workloads.

% \textcolor{red}{TODO: Plot $\rho_{\rm bubble}$ for different batch size for prefiil and decode}

\subsection{Hybrid Parallelism: Balancing Communication and Pipeline Bubbles}

In this section, we provide experiment results on HB. The results illustrate a complementary trade-off between compute and communication of  TP and PP during both prefill and decode. Decoding is dominated by pipeline utilization because only one token is generated per request (i.e., $S=1$), whereas prefill is dominated by computation and communication due to the long input sequence (i.e., $S=\rm ISL$). With HB, we would like to understand how much of this trade-off can be balanced out. Throughout the experiments we assume a fixed number of GPU resources (i.e., $8$ GPUs), and consider the following 4 parallelism configurations: TP8 representing pure tensor parallelism across GPUs, PP8 for pure pipeline parallelism, PP2TP4 for HB of 2 pipeline stages with 4 GPUs in each stage and PP4TP2 representing 4 pipeline stages with 2 GPUs in each stage. 

\begin{figure}[t!]
    \centering
    
    \begin{subfigure}{\textwidth}
        \centering
        \includegraphics[width=\linewidth]{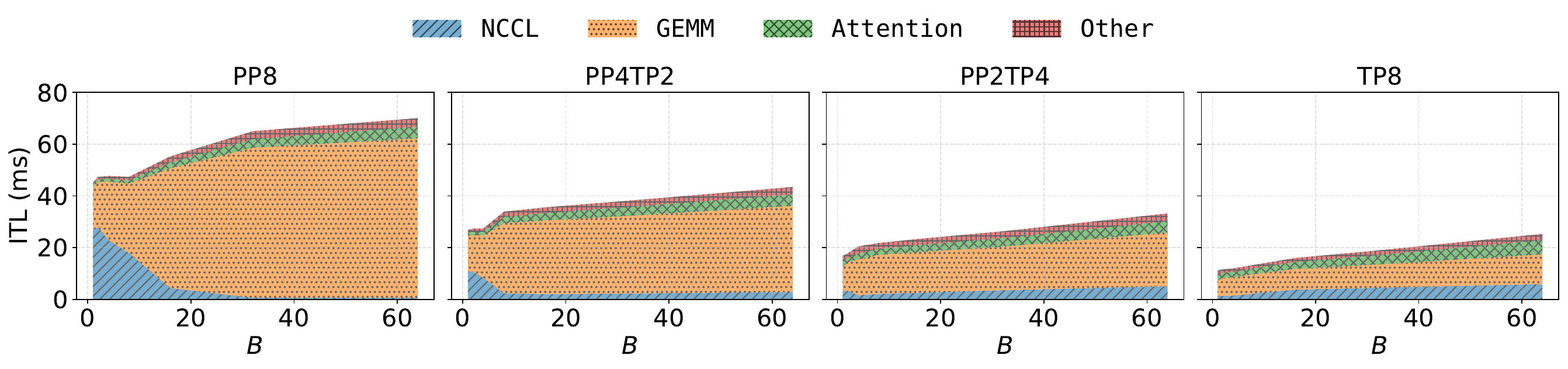}
        \caption{ITL (ms) vs. Batch Size.}    
        \label{fig:HB_ITL}
    \end{subfigure}
    
    % \vspace{0.5cm}
    \begin{subfigure}{\textwidth}
        \centering
        \includegraphics[width=\linewidth]{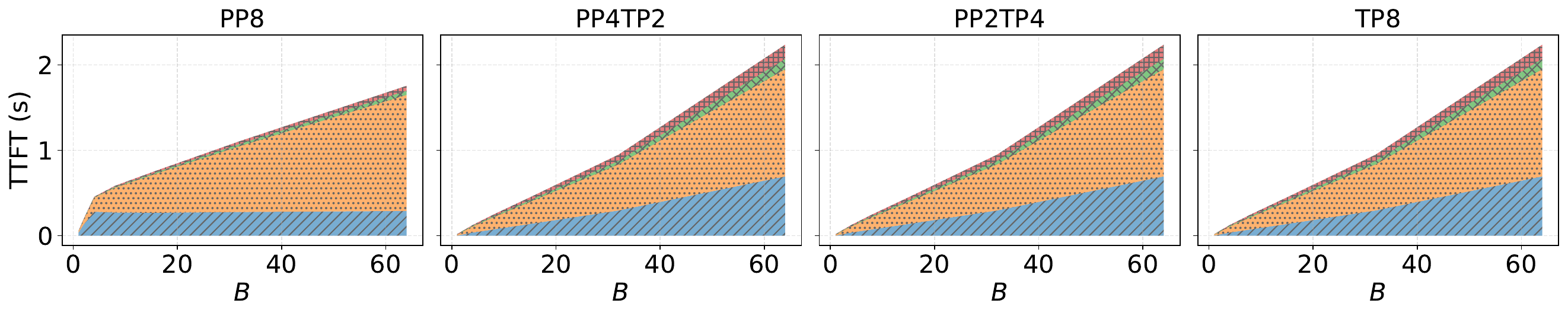}
        \caption{TTFT (s) vs. Batch Size.}
        \label{fig:HB_TTFT}
    \end{subfigure}

    \caption{Kernel-level breakdown of (a) ITL during decoding and (b) TTFT during prefill for hybrid TP-PP configurations for Qwen$2.5$ - $32$B - Instruct using a total of $8$ GPUs. Decode performance is strongly influenced by PP bubble overhead, which diminishes as execution becomes more TP-dominant, whereas prefill latency is primarily determined by TP communication overhead and useful computation.}
    \vspace{-6mm}
\end{figure}

During decoding, as shown in Fig. \ref{fig:HB_ITL}, ITL consistently decreases with increasing TP degree because eliminating pipeline bubbles outweighs the additional TP collective communication. We also observe that the PP-dominant configurations (i.e., PP8 and PP4TP2), especially at small batch sizes, the idle latency contributes a noticeable fraction of the ITL. However, bubble gradually disappears as the decode batch size increases. To explain this, we note that during decoding ($S=1$), the amount of work entering the pipeline is simply $W=B$. Consequently, for small decode batch sizes, the number of work units is insufficient to fully occupy all pipeline stages, resulting in a significant pipeline bubble as shown in \eqref{eq:bubble_frac_HB}. As the batch size increases, the amount of work grows proportionally while the fill/drain overhead remains fixed at $P_{\rm PP} - 1$ stages. Therefore, the bubble fraction rapidly decreases, allowing all pipeline stages to remain continuously occupied, which leads to vanishing pipeline bubble as the decode batch size increases.
Another important observation in Fig. \ref{fig:HB_ITL} is that the bubble becomes progressively smaller as the execution shifts from PP-dominant to TP-dominant configurations (i.e., from PP8 to TP8). This behavior follows directly from \eqref{eq:bubble}, since the bubble latency is proportional to the pipeline depth. Reducing the $P_{\rm PP}$ decreases the number of fill/drain intervals and therefore shortens the idle latency. In the limiting case of TP8 (i.e., $P_{\rm PP}=1$), the pipeline consists of a single stage, eliminating fill/drain entirely, resulting in zero pipeline bubble. Consequently, TP-dominant configurations achieve substantially lower ITL during decoding despite incurring additional TP collective communication.

Interestingly, during prefill, as shown in Fig \ref{fig:HB_TTFT}, TTFT increases as the execution becomes more TP-dominant. The reason is that TP introduces collective communication after distributed matrix multiplications, where latency grows with the message size.
This communication is executed for every transformer layer. Since prefill operates on long sequences ($S=\rm ISL$), the communicating message  size increase linearly with ISL, causing the \verb|NCCL| component to grow rapidly with batch size. In contrast, PP communication involves point-to-point activation transfers between adjacent stages. This makes PP considerably more communication-efficient for large prefill workloads. 

Importantly, we observe in Fig \ref{fig:HB_TTFT} that there is no visible bubble appears in TTFT across different batch sizes, mainly because the behavior of parallelism strategies is fundamentally different from decoding. To explain this we note that during prefill, where all input tokens are processed simultaneously, the amount of work increases to $W=B \times \rm ISL$, which is typically several orders of magnitude larger than during decoding. According to \eqref{eq:bubble_frac_HB}, increasing $W$ dramatically reduces the bubble fraction, for practical input sequence lengths. As a result, the pipeline remains almost fully utilized even for PP-dominant configurations (i.e., PP8 and PP4TP2). Instead, TTFT is almost entirely determined by useful computation and communication rather than pipeline occupancy.

\begin{figure}[t!]
    \centering
    
    \begin{subfigure}[t]{0.49\textwidth}
        \centering
        \includegraphics[width=\linewidth]{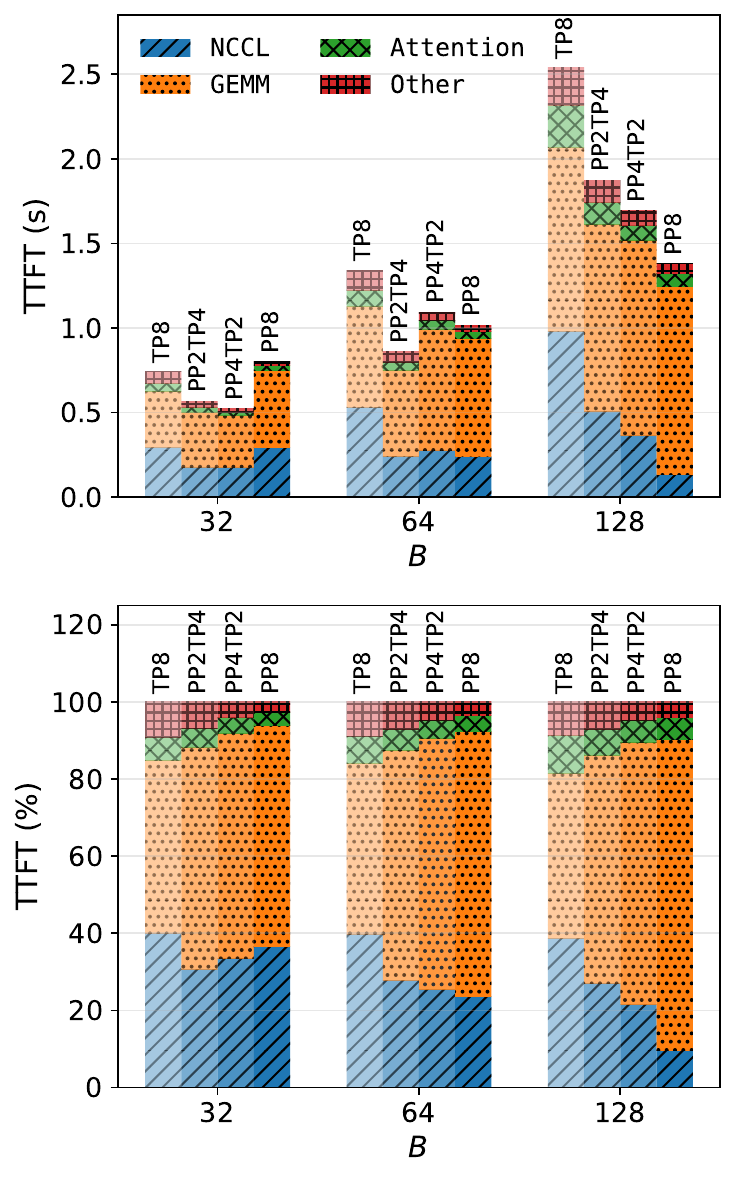}
        \caption{TTFT vs. Batch Size.}
        \label{fig:HB_TTFT_side}
    \end{subfigure}
    \hfill
    \begin{subfigure}[t]{0.49\textwidth}
        \centering
        \includegraphics[width=\linewidth]{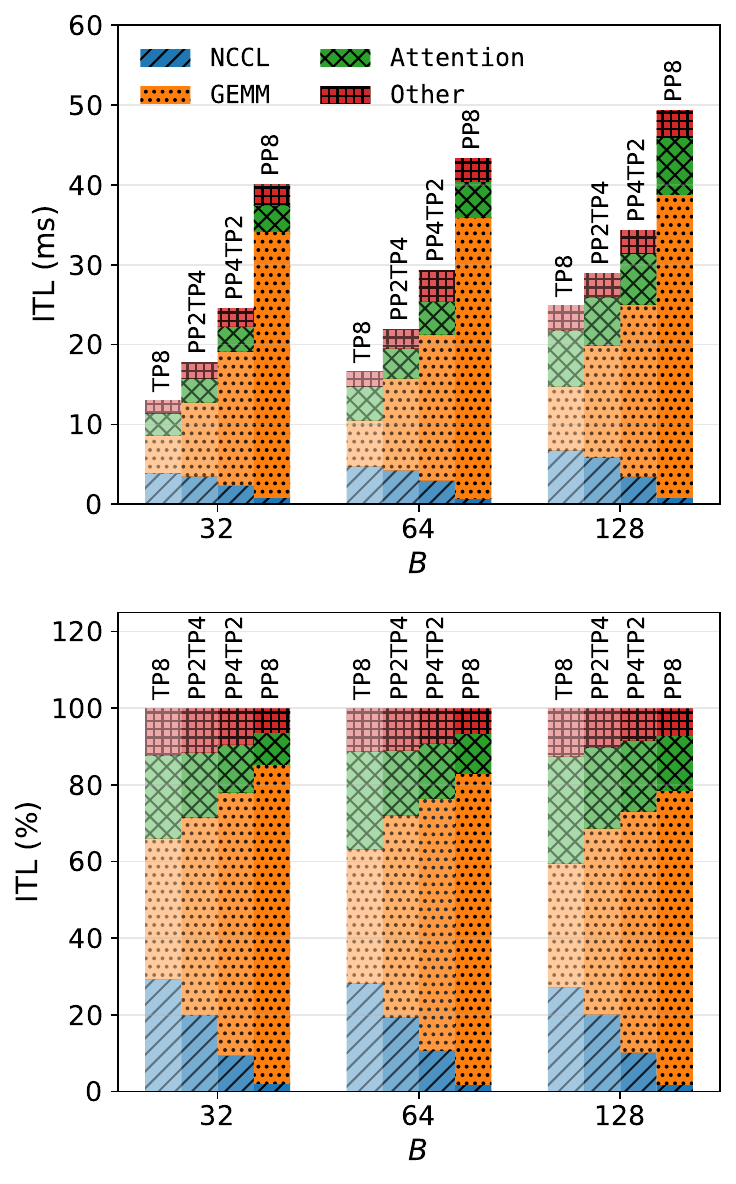}
        \caption{ITL vs. Batch Size.}
        \label{fig:HB_ITL_side}
    \end{subfigure}

    \caption{Trade-off between TP and PP for (a) TTFT during prefill and (b) ITL decode for Qwen$2.5$-$14$B-Instruct across $8$ GPUs. Prefill performance favors PP due to lower communication overhead, while decode performance favors TP due to reduced pipeline bubbles. Hybrid configurations provide intermediate behavior by balancing TP synchronization costs against PP idle-time overhead.}
    \vspace{-5mm}
    \label{fig:HB_kernel_contribution}
\end{figure}

As it is shown in both Fig. \ref{fig:HB_TTFT} and \ref{fig:HB_ITL}, PP-dominant configurations achieve lower TTFT, whereas TP-dominant configurations provide superior ITL during decoding. These observations highlight the inherent trade-off between PP and TP across prefill and decode. To better understand this trade-off, we plot the breakdown of each kernel latency for TTFT and ITL across different batch sizes and different paralleim strategies in Fig. \ref{fig:HB_TTFT_side} and \ref{fig:HB_ITL_side}, respectively. We note that both prefill and decode expose the distinct computational characteristics. Prefill processes the entire input sequence simultaneously, resulting in a large amount of computation per request ($S=\rm ISL$). Consequently, the execution is predominantly compute-bound, and the overall latency is determined by how efficiently the GPUs sustain large matrix multiplications. In contrast, decoding generates only a single token per request ($S=1$), significantly reducing the computation performed in each iteration. As a result, decoding becomes increasingly sensitive to pipeline utilization and scheduling overhead, making pipeline bubbles a dominant performance bottleneck.
In Fig. \ref{fig:HB_TTFT_side}, during prefill, PP-dominant configurations consistently achieve the lowest TTFT at $B=128$ despite using deeper pipelines. This behavior can be explained by the communication characteristics of PP. PP partitions transformer layers across GPUs while keeping the computation of each layer local to a single GPU. Therefore, the only inter-GPU communication is the point-to-point transmission of activations between adjacent stages given in \eqref{eq:PP_comm}, which occurs only once at each pipeline boundary. Since prefill executes long sequences, the computation associated with each stage is substantially larger than the communication cost, allowing communication to be effectively amortized. This observation is consitent with that in Fig. \ref{fig:HB_TTFT}, where the \verb|NCCL| contribution remains relatively small for PP-dominant configurations even as the batch size increases. This indicates that most of the execution time is dedicated to useful \verb|GEMM| rather than communication overhead.

In contrast, TP-dominant configurations exhibit significantly larger TTFT, mainly because of inter-GPU communication after \verb|GEMM| in every layer. Although TP reduces the message size per GPU, every communication requires global synchronization among all TP ranks. The TP communication latency in \eqref{T_NCCL1}, is therefore incurred more frequently throughout the forward pass. Since prefill processes long input sequences ($S=\rm ISL$), the communication volume increases proportionally with $S$, causing the \verb|NCCL| overhead to become substantial. This trend is clearly reflected in Fig. \ref{fig:HB_TTFT_side}, where nearly $40\%$ of the TTFT for TP8 is occupied by \verb|NCCL|, whereas the PP8 configuration spends only a small fraction of its latency on communication. Consequently, despite eliminating pipeline bubbles, TP suffers from excessive synchronization overhead during prefill, making PP the preferred strategy for maximizing TTFT performance.

The decoding phase exhibits the exact opposite behavior. Since only one token is processed per iteration ($S=1$), the useful computation per request becomes extremely small, while the pipeline fill/drain overhead remains unchanged. According to \eqref{eq:bubble_frac_HB}, PP-dominant configurations experience significant idle time at low and moderate batch sizes, which give rise to substantially higher bubble latency. Eliminating these bubbles through TP produces a much larger latency reduction than the additional \verb|NCCL| can offset.
This behavior is clearly shown in Fig. \ref{fig:HB_ITL_side}, where the PP8 configuration exhibits the highest ITL despite having the lowest communication overhead. As the pipeline depth decreases (from PP8 to TP8), the number of fill/drain intervals decreases proportionally, eventually disappearing entirely when $P_{\rm PP} = 1$ (i.e., TP8). Consequently, TP8 achieves the lowest ITL despite the highest \verb|NCCL| among all configurations.

Overall, PP and TP optimize two fundamentally different bottlenecks. PP minimizes synchronization overhead and is therefore ideally suited for the compute-intensive prefill phase, whereas TP minimizes pipeline occupancy loss by eliminating fill/drain overhead and is therefore more effective for latency-critical decoding. The hybrid configurations (PP4TP2 and PP2TP4) naturally provide intermediate behavior, balancing TP communication overhead against PP bubble latency. They offer a compromise between TTFT and ITL rather than optimizing either metric individually. This observation is consistent with the analytical latency model we developed in Section \ref{sec:parallel}, where PP communication, TP communication, and pipeline bubble latency contribute independently to the total inference latency.

\section{Conclusion}

Distributed execution has become essential for serving modern LLMs, yet understanding the performance trade-offs among TP, PP, and HB remains a fundamental challenge. In this paper, we presented a unified analytical framework for modeling distributed LLM inference by decomposing end-to-end latency into computation, inter-GPU communication, and pipeline bubble overhead. We derived analytical models for TP collective communication, PP point-to-point communication, and pipeline utilization, capturing their dependence on batch size, sequence length, hidden dimension, message size, and parallelism degree. Unlike existing empirical studies, our framework provides a principled explanation of how these factors jointly determine inference performance across both the prefill and decoding phases.

Our analysis and experimental evaluation demonstrate that the proposed framework accurately captures the contrasting behavior of different parallelism strategies, explaining why PP is more effective for the compute-intensive prefill phase, whereas TP achieves lower decoding latency by eliminating pipeline bubbles despite higher communication overhead. By providing a unified understanding of the compute, communication and bubble latency trade-offs, the proposed model enables systematic evaluation of parallelism strategies across diverse models, workloads, and hardware configurations. We believe this framework provides a practical foundation for performance modeling, capacity planning, and the design of future distributed LLM serving systems that can adapt parallelism strategies to evolving workload characteristics.

\section*{Acknowledgments} 
The authors would like to acknowledge the support of the Dell leadership team and thank colleagues across Dell for their valuable discussions, constructive feedback, and technical insights that contributed to this research.

\bibliographystyle{ieeetr}  
\bibliography{references}  %%% Remove comment to use the external .bib file (using bibtex).

@misc{nvidia_nccl,
  author       = {{NVIDIA Corporation}},
  title        = {{NVIDIA Collective Communications Library (NCCL)}},
  year         = {2026},
  howpublished = {\url{https://developer.nvidia.com/nccl}},
  note         = {Accessed: 2026-07-10}
}

@misc{nvidia_dynamo,
  author       = {{NVIDIA Corporation}},
  title        = {NVIDIA Dynamo},
  year         = {2026},
  howpublished = {\url{https://www.nvidia.com/en-us/ai/dynamo/}},
  note         = {Accessed: 2026-07-13}
}

@inproceedings{hu2025demystifying,
  title={Demystifying {NCCL}: An in-depth analysis of {GPU} communication protocols and algorithms},
  author={Hu, Z. and Shen, S. and Bonato, T. and Jeaugey, S. and Alexander, C. and Spada, E. and Dinan, J. and Hammond, J. and Hoefler, T.},
  booktitle={2025 IEEE Symp. on High-Performance Interconnects (HOTI)},
  pages={48--59},
  year={2025},
  organization={IEEE}
}

@misc{shoeybi2019megatron,
     title={{Megatron-LM}: Training multi-billion parameter language models using model parallelism},
  author={Shoeybi, M. and Patwary, M. and Puri, R. and LeGresley, P. and Casper, J. and Catanzaro, B.},
  year={2019},
      note = {Available:  {https://arxiv.org/abs/1909.08053}}, 
}

@misc{tensorrtllm_docs,
  author       = {{NVIDIA Corporation}},
  title        = {{TensorRT-LLM} Documentation},
  year         = {2026},
  howpublished = {\url{https://docs.nvidia.com/tensorrt-llm/index.html}},
  note         = {Accessed: 2026-07-13}
}

@misc{vllm_docs,
  author       = {{vLLM Project Contributors}},
  title        = {{vLLM} Documentation},
  year         = {2026},
  howpublished = {\url{https://docs.vllm.ai/en/stable/}},
  note         = {Accessed: 2026-07-13}
}

@inproceedings{kwon2023efficient,
  title={Efficient memory management for large language model serving with paged attention},
  author={Kwon, W. and Li, Z. and Zhuang, S. and Sheng, Y. and Zheng, L. and Yu, C. Hao and Gonzalez, J. and Zhang, H. and Stoica, I.},
  booktitle={Pro. of the 29th symp. on operating systems principles},
  pages={611--626},
  year={2023}
}

@inproceedings{agrawal2024taming,
  title={Taming Throughput-Latency tradeoff in {LLM} inference with Sarathi-Serve},
  author={Agrawal, A. and Kedia, N. and Panwar, A. and Mohan, J. and Kwatra, N. and Gulavani, B. and Tumanov, A. and Ramjee, R.},
  booktitle={18th USENIX symp. on operating sys. design and implementation (OSDI 24)},
  pages={117--134},
  year={2024}
}

@inproceedings{aminabadi2022deepspeed,
  title={Deepspeed-inference: enabling efficient inference of transformer models at unprecedented scale},
  author={Aminabadi, R. and Rajbhandari, S. and Awan, A. A. and Li, C. and Li, D. and Zheng, E. and Ruwase, O. and Smith, S. and Zhang, M. and Rasley, J. and others},
  booktitle={SC22: Int. Conf. for High Performance Computing, Networking, Storage and Analysis},
  pages={1--15},
  year={2022},
  organization={IEEE}
}

@article{narayanan2021efficient,
  title={Efficient large-scale language model training on {GPU} clusters using {Megatron-LM}},
  author={Narayanan, D. and Shoeybi, M. and Casper, J. and LeGresley, P. and Patwary, M. and Korthikanti, V. and Vainbrand, D. and Kashinkunti, P. and Bernauer, J. and Catanzaro, B. and others},
  journal={In Proc. of the Int. conf. for high performance computing, networking, storage and analysis},
  pages={1-15},
  year={2021}
}

@article{huang2019gpipe,
  title={Gpipe: Efficient training of giant neural networks using pipeline parallelism},
  author={Huang, Y. and Cheng, Y. and Bapna, A. and Firat, O. and Chen, D. and Chen, M. and Lee, H. and Ngiam, J. and Le, Q.  and Wu, Y. and others},
  journal={In Proc. Advances neural info. process. sys.},
  volume={32},
  year={2019}
}

@misc{harlap2018pipedream,
      title={Pipedream: Fast and efficient pipeline parallel {DNN} training},
  author={Harlap, A. and Narayanan, D. and Phanishayee, A. and Seshadri, V. and Devanur, N. and Ganger, G. and Gibbons, P.},
  year={2018},
      note = {Available:  {https://arxiv.org/abs/1806.03377}}, 
}

@article{fedus2022switch,
  title={Switch transformers: Scaling to trillion parameter models with simple and efficient sparsity},
  author={Fedus, W. and Zoph, B.and Shazeer, N.},
  journal={Journal of Machine Learning Research},
  volume={23},
  number={120},
  pages={1-39},
  year={2022}
}

@article{10.1016MPP,
author = {Hockney, Roger W.},
title = {The communication challenge for {MPP}: Intel Paragon and Meiko {CS-2}},
year = {1994},
issue_date = {March 1994},
publisher = {Elsevier Science Publishers B. V.},
address = {NLD},
volume = {20},
number = {3},
issn = {0167-8191},
url = {https://doi.org/10.1016/S0167-8191(06)80021-9},
doi = {10.1016/S0167-8191(06)80021-9},
journal = {Parallel Comput.},
month = mar,
pages = {389–398},
numpages = {10}
}

@INPROCEEDINGS{647225,
  author={Iannello, G. and Lauria, M. and Mercolino, S.},
  booktitle={Proc. Sixth Euromicro Workshop on Parallel and Distributed Process.}, 
  title={{LogP} performance characterization of fast messages atop Myrinet}, 
  year={1998},
  volume={},
  number={},
  pages={395-401},
  doi={10.1109/EMPDP.1998.647225}}

@article{10.1145/1498765.1498785,
author = {Williams, S. and Waterman, A. and Patterson, D.},
title = {Roofline: an insightful visual performance model for multicore architectures},
year = {2009},
issue_date = {April 2009},
publisher = {Association for Computing Machinery},
address = {New York, NY, USA},
volume = {52},
number = {4},
issn = {0001-0782},
url = {https://doi.org/10.1145/1498765.1498785},
doi = {10.1145/1498765.1498785},
journal = {Commun. ACM},
month = apr,
pages = {65–76},
numpages = {12}
}

@article{zhong2024distserve,
  title={{DistServe}: Disaggregating prefill and decoding for goodput-optimized large language model serving},
  author={Zhong, Y. and Liu, S. and Chen, J. and Hu, J. and Zhu, Y. and Liu, X. and Jin, X. and Zhang, H.},
  journal={18th USENIX Symp. Operating Sys. Design and Implementation (OSDI 24)},
  pages={193--210},
  year={2024}
}

@inproceedings{patel2024splitwise,
  title={Splitwise: Efficient generative llm inference using phase splitting},
  author={Patel, P. and Choukse, E. and Zhang, C. and Shah, A. and Goiri, {\'I}. and Maleki, S. and Bianchini, R.},
  booktitle={2024 ACM/IEEE 51st Annual Int. Symp. on Computer Architecture (ISCA)},
  pages={118--132},
  year={2024},
  organization={IEEE}
}

@article{qin2024mooncake,
  title={Mooncake: A {KV} cache-centric disaggregated architecture for {LLM} serving},
  author={Qin, R. and Li, Z. and He, W. and Cui, J. and Tang, H. and Ren, F. and Ma, T. and Cai, S. and Zhang, Y. and Zhang, M. and others},
  journal={ACM Transactions on Storage},
  year={2024},
  publisher={ACM New York, NY}
}

@misc{qwen2025qwen25technicalreport,
      title={Qwen2.5 Technical Report}, 
      author={A. Yang and B. Yang and B. Zhang and B. Hui and B. Zheng and B. Yu and C. Li and D. Liu and F. Huang and H. Wei and H. Lin and J. Yang and J. Tu and J. Zhang and J. Yang and J. Yang and J. Zhou and J. Lin and K. Dang and K. Lu and K. Bao and K. Yang and L. Yu and M. Li and M. Xue and P. Zhang and Q. Zhu and R. Men and R. Lin and T. Li and T. Tang and T. Xia and X. Ren and X. Ren and Y. Fan and Y. Su and Y. Zhang and Y. Wan and Y. Liu and Z. Cui and Z. Zhang and Z. Qiu},
      year={2025},
      eprint={2412.15115},
      archivePrefix={arXiv},
      primaryClass={cs.CL},
      note = {Available: {https://arxiv.org/abs/2412.15115}}, 
}

@misc{NvdaA100,
  author       = {{NVIDIA}},
  title        = {{NVIDIA} {A100} Tensor Core {GPU}: Unprecedented Acceleration at Every Scale },
  year         = {2026},
  howpublished = {\url{https://www.nvidia.com/content/dam/en-zz/Solutions/Data-Center/a100/pdf/nvidia-a100-datasheet-nvidia-us-2188504-web.pdf}},
  note         = {Accessed: 2026-09-25}
}

@misc{vllm15,
  author       = {{vLLM v0.15.0}},
  title        = {{vLLM}: Easy, fast, and cheap LLM serving for everyone},
  year         = {2026},
  howpublished = {\url{https://docs.vllm.ai/en/v0.15.0/}},
  note         = {Accessed: 2026-09-25}
}

@misc{Nsight_sys_v2026_5,
  title        = {{NVIDIA Nsight Systems v2026.5.1 }},
  year         = {2026},
  howpublished = {\url{https://developer.nvidia.com/nsight-systems}},
  note         = {Accessed: 2026-09-25}
}


% George Kour and Raid Saabne.
% \newblock Real-time segmentation of on-line handwritten arabic script.
% \newblock In {\em Frontiers in Handwriting Recognition (ICFHR), 2014 14th
%   International Conference on}, pages 417--422. IEEE, 2014.

% \end{thebibliography}

% \newpage
% \appendix  % Start appendix
% \section*{Appendix}

% \section{Tensor Parallelism: ITL and TTFT vs Batch Size}

\end{document}